\documentclass[subeqn,11pt, a4paper]{amsart}

\usepackage[margin=35mm]{geometry}

\usepackage{amssymb,amsmath,amsfonts,amsthm, amscd, mathrsfs, helvet, mathtools}
\usepackage{bm, stmaryrd}
\usepackage{graphicx, verbatim}
\usepackage[mathcal]{euscript}
\usepackage{color}
\usepackage[all]{xy}
\usepackage{relsize}

\usepackage{tikz}
\usetikzlibrary{cd}
\usetikzlibrary{matrix,calc}
\usetikzlibrary{decorations.markings,shapes.geometric,shapes.misc}
\usetikzlibrary{decorations.pathreplacing,positioning}
\usetikzlibrary{arrows}

\tikzset{cross/.style={cross out, draw=black, minimum size=2*(#1-\pgflinewidth), inner sep=0pt, outer sep=0pt}, cross/.default={1pt}}

\tikzset{cross/.style={cross out, draw=black, minimum size=2*(#1-\pgflinewidth), inner sep=0pt, outer sep=0pt}, cross/.default={1pt}}

\allowdisplaybreaks[1]

\usepackage{pict2e}

\usepackage{hyperref}
\hypersetup{
     colorlinks=false,         
     linkcolor=darkred,
     citecolor=blue,
}

\makeatletter
\DeclareFontFamily{OMX}{MnSymbolE}{}
\DeclareSymbolFont{MnLargeSymbols}{OMX}{MnSymbolE}{m}{n}
\SetSymbolFont{MnLargeSymbols}{bold}{OMX}{MnSymbolE}{b}{n}
\DeclareFontShape{OMX}{MnSymbolE}{m}{n}{
    <-6>  MnSymbolE5
   <6-7>  MnSymbolE6
   <7-8>  MnSymbolE7
   <8-9>  MnSymbolE8
   <9-10> MnSymbolE9
  <10-12> MnSymbolE10
  <12->   MnSymbolE12
}{}
\DeclareFontShape{OMX}{MnSymbolE}{b}{n}{
    <-6>  MnSymbolE-Bold5
   <6-7>  MnSymbolE-Bold6
   <7-8>  MnSymbolE-Bold7
   <8-9>  MnSymbolE-Bold8
   <9-10> MnSymbolE-Bold9
  <10-12> MnSymbolE-Bold10
  <12->   MnSymbolE-Bold12
}{}

\let\llangle\@undefined
\let\rrangle\@undefined
\DeclareMathDelimiter{\llangle}{\mathopen}%
                     {MnLargeSymbols}{'164}{MnLargeSymbols}{'164}
\DeclareMathDelimiter{\rrangle}{\mathclose}%
                     {MnLargeSymbols}{'171}{MnLargeSymbols}{'171}
\makeatother



\definecolor{myPurple}{rgb}{0.5,0.1,0.6}
\definecolor{myOrange}{rgb}{1.0,0.5,0.0}
\definecolor{myRed}{rgb}{1.0,0.0,0.0}
\definecolor{myGreen}{rgb}{0.0,0.5,0.0}
\definecolor{LatexBlue}{rgb}{0.211765,0.227451,0.666667}
\definecolor{myBlue}{rgb}{0.0,0.0,1.0}
\definecolor{myBlack}{rgb}{0.0,0.0,0.0}
\definecolor{myGray}{rgb}{0.3,0.3,0.3}

\theoremstyle{plain}

\newtheorem*{theorem*}{Theorem}

\newtheorem*{proposition*}{Proposition}

\theoremstyle{definition}

\DeclareMathOperator{\res}{res}

\DeclareMathOperator{\Omegaone}{\Omega^{(1)}}
\DeclareMathOperator{\Tr}{Tr}

\newcommand{\CP}{\mathbb{P}^1}

\def\gl{\mathfrak{gl}}

\def\L{\mathcal{L}}

\def\CS{\mathrm{CS}}

\newcommand{\lau}[1]{(\kern-.2em( #1 )\kern-.2em)}

\newcommand{\ms}[1]{\mathsf{#1}}
\newcommand{\cpb}[2]{\{\! | #1, #2| \! \}}
\newcommand{\ip}[2]{#1\lrcorner #2}
\newcommand{\parder}[2]{\frac{\partial #1}{\partial #2}}
\newcommand{\ie}{{\it i.e.}\ }

\def\be{\begin{equation}}
\def\ee{\end{equation}}
\def\bea{\begin{eqnarray}}
\def\eea{\end{eqnarray}}

\def\CC{\mathbb{C}}
\def\CP{\mathbb{C}P^1}

\def\RR{\mathbb{R}}

\def\ZZ{\mathbb{Z}}

\def\ii{{\rm i}}

\def\1{\bm{1}}

\numberwithin{equation}{section}

\DeclareUnicodeCharacter{2212}{-}

\begin{document}

\title[Zakharov-Mikhailov action: $4$d CS origin and covariant Lax algebra]{On the Zakharov-Mikhailov action:\\
$4$d Chern-Simons origin and covariant\\
Poisson algebra of the Lax connection}

\author{Vincent Caudrelier}
\address{School of Mathematics, University of Leeds, LS2 9JT, U.K.}\email{v.caudrelier@leeds.ac.uk}
\author{Matteo Stoppato (Corresponding author)}
\address{School of Mathematics, University of Leeds, LS2 9JT, U.K.}\email{mmms@leeds.ac.uk} 
\author{Beno\^{\i}t Vicedo}
\address{Department of Mathematics, University of York, York YO10 5DD, U.K.} \email{benoit.vicedo@gmail.com}

\begin{abstract}
We derive the $2$d Zakharov-Mikhailov action from $4$d Chern-Simons theory. This $2$d action is known to produce as equations of motion the flatness condition of a large class of Lax connections of Zakharov-Shabat type, which includes an ultralocal variant of the principal chiral model as a special case. At the $2$d level, we determine for the first time the covariant Poisson bracket $r$-matrix structure of the Zakharov-Shabat Lax connection, which is of rational type. The flatness condition is then derived as a covariant Hamilton equation. We obtain a remarkable formula for the covariant Hamiltonian in term of the Lax connection which is the covariant analogue of the well-known formula ``$H=\Tr L^2$''.
\end{abstract}

\maketitle


\input{epsf}

\section{Introduction}

Classical integrable field theories in two dimensions are characterised by the existence of a Lax connection which is on-shell flat and depends meromorphically on an auxiliary complex curve, typically the Riemann sphere. Determining whether a given field theory is integrable or not is usually a very difficult problem since there is no systematic way of constructing a suitable Lax connection, if one exists.

Over the past couple of years, however, two closely related general frameworks have emerged for explaining the algebraic and geometric origins of Lax connections in $2$d integrable field theories. From a Hamiltonian perspective, an origin was proposed in \cite{Vicedo:2017cge}, and further developed in \cite{Delduc:2019bcl}, based on Gaudin models associated with untwisted affine Kac-Moody algebras and the representation theory of such algebras. From a Lagrangian perspective, an origin was proposed by Costello and Yamazaki in \cite{Costello:2019tri}, following earlier work on integrable spin chains in \cite{Costello:2013zra, Costello:2013sla, Witten:2016spx, Costello:2017dso, Costello:2018gyb}, based on the introduction of surface defects in $4$d Chern-Simons theory. A much older connection between Lagrangians for (hierarchies of) integrable field theories in $2$d and field theories of Wess-Zumino-Witten type was pioneered in \cite{N}.

\medskip

In the Hamiltonian formulation of integrable field theories, there is an important dichotomy between so-called `ultralocal' and `non-ultralocal' theories. This distinction is based on the classical $r$-matrix formalism \cite{Skly1,Skly2}, specifically on whether or not the $r$-matrix of the given integrable field theory is skew-symmetric \cite{SemenovTianShansky:1983ik, Maillet:1985fn, Maillet:1985ek}.

The affine Gaudin model perspective on integrable field theories was specifically developed in \cite{Vicedo:2017cge} to address the problem of quantisation of non-ultralocal theories. Note that a related approach was used in \cite{Zotov:2010kb} to treat ultralocal field theories as Gaudin models associated with affine Kac-Moody algebras. On the other hand, it was demonstrated on examples in \cite{Costello:2019tri} that both ultralocal and non-ultralocal field theories can be described from the perspective of $4$d Chern-Simons theory. In the non-ultralocal case, further examples were shown in \cite{Delduc:2019whp} to fit within this framework, and more recently a very general family of new non-ultralocal integrable field theories was also constructed using this approach in \cite{Lacroix:2020flf} following \cite{Benini:2020skc}. By performing a Hamiltonian analysis of $4$d Chern-Simons theory, it was shown in \cite{Vicedo:2019dej} that in the case of non-ultralocal field theories this frameworks is, in fact, intimately related to that of affine Gaudin models. By contrast, a Hamiltonian analysis of the class of ultralocal theories from the perspective of $4$d Chern-Simons theory has so far not been considered. The main purpose of this paper is to initiate such a study.

\medskip

In fact, very recently, an independent line of research emerged in \cite{CS1} where the classical $r$-matrix structure was derived in the context of covariant Hamiltonian field theory. In that setting, a covariant Poisson bracket replaces the standard Poisson bracket and the $r$-matrix determines the Poisson algebra satisfied by the whole Lax connection (a $1$-form) and not just by its spatial component, called the Lax matrix.

Such results have been established successfully for ultralocal theories with rational $r$-matrix (nonlinear Schr\"odinger and modified Korteweg-de Vries) and trigonometric $r$-matrix (sine-Gordon). However, the generalisation to non-ultralocal theories has resisted all attempts so far. In particular, the famous example of the principal chiral model, which is intrinsically non-ultralocal, does not seem to be easily amenable to this covariant formalism. Nevertheless, a certain reduction of the principal chiral model dynamics can be reproduced by an ultralocal integrable field theory \cite{Faddeev:1985qu}, for which an action was obtained in \cite{Appadu:2017fff}. This model can be seen as a special case of a large class of models with Lax pairs of Zakharov-Shabat type which derive from an action first introduced by Zakharov and Mikhailov \cite{ZakhMikh}. Our observation is that this general class of models admits an ultralocal $r$-matrix structure of rational type and is therefore suited for a covariant Hamiltonian treatment.

\medskip

The main goal of the present work is to begin exploring the {\it covariant} Hamiltonian structure of certain ultralocal integrable field theories which can be obtained from the $4$d Chern-Simons perspective, using the Zakharov-Mikhailov class of models as our guiding example. The covariant approach to integrable field theories initiated in \cite{CS1} is in contrast with the long tradition of analysing the standard Hamiltonian formulation of integrable field theories and may offer new insights when it comes to their (covariant) quantisation. An interesting by-product of our approach is the interpretation of the flatness condition of the Lax connection as a covariant Hamilton equation associated with a covariant Hamiltonian which we derive from the Zakharov-Mikhailov action. 

\medskip

In Section \ref{CS-ZM}, we show that the Zakharov-Mikhailov action of \cite{ZakhMikh} can be derived from $4$d Chern-Simons theory.
Since, in our case, the meromorphic $1$-form $\omega$ appearing in the $4$d Chern-Simons action is $\omega = dz$, it has a double pole at infinity so we follow a similar approach to \cite{Benini:2020skc} by first regularising the action of $4$d Chern-Simons theory. We then couple minimally the $4$d gauge field $A$ to a collection of Lie group valued fields $\{ \phi_m \}_{m=1}^{N_1}$ and $\{ \psi_n \}_{n=1}^{N_2}$ localised along surface defects.

In Section \ref{sec: covariant}, we derive the covariant Poisson algebra satisfied by the Lax connection of the Zakharov-Mikhailov class of models. We also present the covariant Hamiltonian of the theory and derive a remarkable formula connecting it to the Lax connection. This formula represents the covariant analogue of the well-known formula relating a traditional Hamiltonian $H$ with the Lax matrix $L$ which we write schematically as ``$H= \Tr L^2$''. We show that the flatness condition of the Lax connection takes the form of a covariant Hamilton equation, thus giving it a new interpretation in this context. The results of this section rely heavily on the variational bicomplex formalism as presented in \cite[Chap. 19]{Dickey} and on ideas developed for instance in \cite{Kanat}. For a detailed account geared specifically towards the implementation of these ideas in $2$d integrable field theories, we refer the reader to \cite{CS1}.

\section{Zakharov-Mikhailov action from $4$d Chern-Simons}\label{CS-ZM}

Using the same notation as in \cite{ZakhMikh}, we let $N_1, N_2 \in \ZZ_{\geq 1}$ and fix subsets $\{ a_m \}_{m=1}^{N_1}$ and $\{ b_n \}_{n=1}^{N_2}$ of $\CP$, which we take to be \emph{disjoint} as in \cite{ZakhMikh}, namely $a_m \neq b_n$ for all $m = 1, \ldots, N_1$ and $n = 1, \ldots, N_2$.
We parametrise the plane $\Sigma \coloneqq \RR^2$ with `light-cone' coordinates $\eta$ and $\xi$.

We shall work with the general linear group $GL_N$ with Lie algebra $\gl_N$ of $N \times N$ matrices, following \cite{ZakhMikh}, but we expect our results to hold more generally for any semisimple Lie algebra. We denote the trace by $\Tr : \gl_N \to \RR$.

Let $X \coloneqq \Sigma \times \CP$. We shall use the notation
\begin{align*}
\Tr \bigg( \sum_A \ms u_A dx^A \wedge \sum_B \ms v_B dx^B \bigg) \coloneqq \sum_{A,B} \Tr(\ms u_A \ms v_B) dx^A \wedge dx^B
\end{align*}
for $\gl_N$-valued $p$- and $q$-forms on $X$, where $p, q = 0, \ldots, 4$, $\ms u_A, \ms v_B \in \gl_N$ and $A,B$ are multi-indices with $|A| = p$ and $|B| = q$ so that $\{ dx^A \}$ and $\{ dx^B \}$ denote bases of $1$-forms for the space $\Omega^p(X)$ and $\Omega^q(X)$, respectively.

\subsection{Regularised $4$d Chern-Simons action} \label{sec: reg 4d action}

Since the $2$d integrable field theory we want to describe is ultralocal, we consider the meromorphic $1$-form $\omega = dz$. The Lagrangian of the corresponding $4$d Chern-Simons theory is given by
\begin{equation} \label{Lagrangian 4dCS}
L_{\rm CS} \coloneqq \frac{\ii}{4 \pi} dz \wedge \CS(A),
\end{equation}
where $\CS(A) \coloneqq \Tr (A \wedge dA + \tfrac 23 A \wedge A \wedge A )$ denotes the Chern-Simons $3$-form and $A$ is a $\gl_N$-valued $1$-form on $X$ which we can decompose as
\begin{equation} \label{A decomp}
A = A_\xi d\xi + A_\eta d\eta + A_{\bar z} d\bar z.
\end{equation}
Note that there is no need to include a $dz$-component since this would drop out from the Lagrangian \eqref{Lagrangian 4dCS}.
The components $A_\xi$, $A_\eta$ and $A_{\bar z}$ are taken to be smooth functions away from the set of marked points $\{ a_m \}_{m=1}^{N_1}$ and $\{ b_n \}_{n=1}^{N_2}$, but it will be important for later to allow $A_\xi$ and $A_\eta$ to be singular at those points. Specifically, we will assume that these components can be written locally as $A_\xi = (z - a_m)^{-1} B_{m, \xi}$ near $a_m$ for $m = 1, \ldots, N_1$ and as $A_\eta = (z - b_n)^{-1} B_{n, \eta}$ near $b_n$ for $n = 1, \ldots, N_2$, where $B_{m, \xi}$ and $B_{n, \eta}$ are smooth functions on $X$. One easily checks that, despite the presence of these singularities, the Lagrangian \eqref{Lagrangian 4dCS} remains locally integrable near $\Sigma \times \{ a_m \}$ for $m = 1, \ldots, N_1$ and near $\Sigma \times \{ b_n \}$ for $n = 1, \ldots, N_2$.

However, since the $1$-form $dz$ has a double pole at $z = \infty$, the $4$-form $dz \wedge \CS(A)$ is not locally integrable near $\Sigma \times \{ \infty \}$. For this reason we need to regularise the action which we do following \cite{Benini:2020skc}. First, note that
\begin{align*}
d\CS(A) &= \Tr (d A \wedge dA + \tfrac 23 dA \wedge A \wedge A - \tfrac 23 A \wedge dA \wedge A + \tfrac 23 A \wedge A \wedge dA )\\
&= \Tr (F(A) \wedge F(A) )
\end{align*}
where $F(A) \coloneqq dA + A \wedge A \in \Omega^2(\Sigma \times \CP, \gl_N)$ is the curvature of $A$. Here we have used the fact that $\Tr (A \wedge A \wedge A \wedge A) = 0$ for any $1$-form $A \in \Omega^1(X, \gl_N)$ by the cyclicity of the trace.

We can now rewrite the Lagrangian \eqref{Lagrangian 4dCS} of $4$d Chern-Simons theory as
\begin{equation*}
L_{\rm CS} = \frac{\ii}{4 \pi} d \big( z \, \CS(A) \big) - \frac{\ii}{4 \pi} z \, \Tr (F(A) \wedge F(A) ),
\end{equation*}
where the first term is exact but has a double pole at infinity while the second term only has a simple pole and is therefore locally integrable near $\Sigma \times \{ \infty \}$. We therefore define the regularised action of $4$d Chern-Simons theory by dropping the exact term above and keeping only the second term, namely we set
\begin{equation} \label{4d CS action}
S_{4d}(A) \coloneqq - \frac{\ii}{4 \pi} \int_X z \Tr ( F(A) \wedge F(A) ).
\end{equation}
Note that we can continue to assume that $A$ has no $dz$-component, namely it can be expressed as in \eqref{A decomp}, since \eqref{4d CS action} is invariant under local transformations
\begin{equation} \label{gauge tr chi}
A \mapsto A + \chi dz
\end{equation}
for any $\chi \in C^\infty(X, \gl_N)$. Indeed, under such a transformation the curvature $F(A)$ transforms as $F(A) \mapsto F(A) + (d \chi + [A, \chi]) \wedge dz$ from which it follows that
\begin{align*}
z \Tr(F(A) \wedge F(A)) &\longmapsto z \Tr(F(A) \wedge F(A)) + 2 z \Tr\big( F(A) \wedge (d \chi + [A, \chi]) \big) \wedge dz\\
&\qquad = z \Tr(F(A) \wedge F(A)) + 2 \, d \big( z dz \wedge \Tr(F(A) \chi) \big)
\end{align*}
where in the second line we have used the fact that $dF(A) = F(A) \wedge A - A \wedge F(A)$.

The action \eqref{4d CS action} is also invariant under gauge transformations
\begin{equation} \label{gauge transformation}
A \longmapsto \null^g A \coloneqq - dg g^{-1} + g A g^{-1}
\end{equation}
for any $g \in C^\infty(X, G)$. Indeed, the transformation of the curvature $F(A)$ under a gauge transformation \eqref{gauge transformation} is given by conjugation $F(\null^g A) = g F(A) g^{-1}$, hence the result follows by the invariance of the trace.

\subsection{Adding surface defects}

We would like to modify the action \eqref{4d CS action} by adding to it terms which couple the $4$d bulk field $A$ to a collection of $2$d fields localised on the surface defects $\Sigma \times \{ a_m \}$ and $\Sigma \times \{ b_n \}$ for $m=1, \ldots, N_1$ and $n = 1, \ldots, N_2$. We shall make use of the embedding $\iota_x : \Sigma \times \{ x \} \hookrightarrow X$ for $x \in \{ a_m \}_{m=1}^{N_1} \cup \{ b_n \}_{n=1}^{N_2}$.

Specifically, to each marked point $a_m$ for $m = 1, \ldots, N_1$ we associate a Lie group valued field $\phi_m \in C^\infty(\Sigma, GL_N)$ which we think of as living on the surface defect $\Sigma \times \{ a_m \}$.
Likewise, to each of the marked points $b_n$ for $n = 1, \ldots, N_2$ we associate a Lie group valued field $\psi_n \in C^\infty(\Sigma, GL_N)$, living on the surface defect $\Sigma \times \{ b_n \}$. Let us also fix constant non-dynamical elements $U^{(0)}_m, V^{(0)}_n$ in $\gl_N$ for $m = 1, \ldots, N_1$ and $n = 1, \ldots, N_2$. Note that the $2$d fields $\phi_m$ and $\psi_n$ are effectively valued in a quotient of $GL_N$ by the stabilisers of $U^{(0)}_m$ and $V^{(0)}_n$, respectively.

Following the discussion of order defects in \cite{Costello:2019tri}, we now couple the $4$d gauge field $A$ to the collection of $2$d fields $\phi_m$ and $\psi_n$ on the surface defects by replacing the regularised $4$d Chern-Simons action \eqref{4d CS action} with
\begin{align} \label{defect 4d action}
S\big( A, \{ \phi_m \}_{m=1}^{N_1}, \{ \psi_n \}_{n=1}^{N_2} \big) \coloneqq S_{\rm 4d}(A) + S_{\rm defect}\big( A, \{ \phi_m \}_{m=1}^{N_1}, \{ \psi_n \}_{n=1}^{N_2} \big)
\end{align}
where we define
\begin{align} \label{defect 2d action}
S_{\rm defect}\big( A, \{ \phi_m \}_{m=1}^{N_1}, \{ \psi_n \}_{n=1}^{N_2} \big) \coloneqq &- \sum_{m=1}^{N_1} \int_{\Sigma \times \{ a_m \}} \Tr \big( \phi_m^{-1} (d_\Sigma + \iota_{a_m}^\ast A) \phi_m U^{(0)}_m \big) \wedge d\xi \notag\\
&\; - \sum_{n=1}^{N_2} \int_{\Sigma \times \{ b_n \}} \Tr \big( \psi_n^{-1} (d_\Sigma + \iota_{b_n}^\ast A) \psi_n V^{(0)}_n \big) \wedge d\eta.
\end{align}
Here $d_\Sigma$ denotes the de Rham differential on $\Sigma$.

To maintain the gauge invariance of the action under \eqref{gauge transformation} after introducing the surface defects, we need to let the $2$d fields transform as
\begin{equation} \label{gauge transf phi psi}
\phi_m \longmapsto g \phi_m, \qquad
\psi_n \longmapsto g \psi_n.
\end{equation}
It is straightforward to check that the extended action \eqref{defect 4d action} is then gauge invariant since the expressions $\phi_m^{-1} (d_\Sigma + \iota_{a_m}^\ast A) \phi_m$ and $\psi_n^{-1} (d_\Sigma + \iota_{b_n}^\ast A) \psi_n$ are themselves gauge invariant.

\subsection{Bulk equations of motion}

Consider the variation $A \mapsto A + \epsilon a$ of the action \eqref{defect 4d action}, for some arbitrary $a = a_\eta d\eta + a_\xi d\xi + a_{\bar z} d\bar z \in \Omega^1_c(X, \gl_N)$ of compact support. This reads
\begin{align*}
&\delta_a S\big( A, \{ \phi_m \}_{m=1}^{N_1}, \{ \psi_n \}_{n=1}^{N_2} \big) \coloneqq \frac{d}{d\epsilon}\bigg|_{\epsilon = 0} S\big( A + \epsilon a, \{ \phi_m \}_{m=1}^{N_1}, \{ \psi_n \}_{n=1}^{N_2} \big) \\
&\qquad\qquad = \frac{\ii}{2 \pi} \int_X dz \wedge \Tr(a \wedge F(A) ) - \sum_{m=1}^{N_1} \int_{\Sigma \times \{ a_m \}} \Tr(a_\eta U_m) d\eta \wedge d\xi\\
&\qquad\qquad\qquad\qquad\qquad\qquad\qquad\qquad\; + \sum_{n=1}^{N_2} \int_{\Sigma \times \{ b_n \}} \Tr (a_\xi V_n) d\eta \wedge d\xi,
\end{align*}
where we introduced $U_m \coloneqq \phi_m U^{(0)}_m \phi_m^{-1}$ for all $m = 1, \ldots, N_1$ and $V_n \coloneqq \psi_n V^{(0)}_n \psi_n^{-1}$ for all $n = 1, \ldots, N_2$. As we will see below, the 2d action obtained from our 4d action with defects effectively gives equations of motion for $U_m$ and $V_n$ which are valued in the Lie algebra $\gl_N$. Without any particular model in mind, it is a matter of taste at this stage whether one wants to interpret the fields of the theory as being those Lie algebra elements or the group elements $\phi_m$ and $\psi_n$. In the former interpretation, the phase space is thus the (co)adjoint orbit through $U^{(0)}_m$ and $V^{(0)}_n$.

In the first term on the right hand side above we have dropped a boundary term which vanishes since $a \in \Omega^1_c(X, \gl_N)$ is of compact support.
The curvature $F(A)$ is given in components by (recall that \eqref{gauge tr chi} ensures that we can take $A$ with no $dz$-component)
\begin{align*}
F(A) &= \big( \partial_\eta A_\xi - \partial_\xi A_\eta + [A_\eta, A_\xi] \big) d\eta \wedge d\xi\\
&\qquad + \big( \partial_{\bar z} A_\xi - \partial_\xi A_{\bar z} + [A_{\bar z}, A_\xi] \big) d\bar z \wedge d\xi\\
&\qquad\qquad + \big( \partial_{\bar z} A_\eta - \partial_\eta A_{\bar z} + [A_{\bar z}, A_\eta] \big) d\bar z \wedge d\eta+dz\wedge \partial_z A\,.
\end{align*}
Note that the last term does not contribute to the equation of motion.
The $A_{\bar z}$ equation of motion is then given by
\begin{equation} \label{A flatness}
\partial_\eta A_\xi - \partial_\xi A_\eta + [A_\eta, A_\xi] = 0,
\end{equation}
which will become the zero curvature equation for the Lax connection. On the other hand, the $A_\eta$ and $A_\xi$ equations of motion respectively read
\begin{subequations} \label{eta xi eom}
\begin{align}
\partial_{\bar z} A_\xi - \partial_\xi A_{\bar z} + [A_{\bar z}, A_\xi] &= 2 \pi \ii \sum_{m=1}^{N_1} U_m \delta(z - a_m),\\
\partial_{\bar z} A_\eta - \partial_\eta A_{\bar z} + [A_{\bar z}, A_\eta] &= 2 \pi \ii \sum_{n=1}^{N_2} V_n \delta(z - b_n)
\end{align}
\end{subequations}
where the $\delta$-functions, satisfying the property
\begin{equation} \label{delta property}
\int_{\CP} f(\xi, \eta, z) \delta(z- x) dz \wedge d\bar z = f(\xi, \eta, x)
\end{equation}
for any $x \in \mathbb{C}$ and any smooth function $f$ on $X$, come from the fact that the surface defect terms are localised at $z = a_m$ or $z = b_n$.

\subsection{Lax connection} \label{sec: Lax connection}

Given the resemblance of \eqref{A flatness} with the zero curvature equation satisfied by the Lax connection, we would like to turn $A$ into the Lax connection itself. There are two obvious issues with this.

The first main issue is that $A$ has an additional $d\bar z$-component compared to the Lax connection $\L = \L_\eta d\eta + \L_\xi d\xi$. We can eliminate this problem by focusing on field configurations with no $d \bar z$-component. This will break some of the gauge invariance since we must now impose that \eqref{gauge transformation} does not re-create any $d\bar z$-component in the gauge field. In other words, we impose that
\begin{equation} \label{A bar z gauge}
A_{\bar z} = 0, \qquad \bar\partial g g^{-1} = 0.
\end{equation}
An obvious way to ensure the latter condition is to take $g \in C^\infty(\Sigma, GL_N)$, \emph{i.e.} $g$ no longer depends on $\CP$. These residual gauge transformations will correspond to gauge transformations in the $2$d theory.

The next difference between $A = A_\eta d\eta + A_\xi d\xi$ and a Lax connection is that the former depends smoothly on $\CP$, with singularities at the marked points $a_m$ and $b_n$ of the form described in \S\ref{sec: reg 4d action}, while the latter is meromorphic on $\CP$. This issue is resolved by focusing again on a subset of gauge fields which satisfy the equations of motion \eqref{eta xi eom}. Having fixed $A_{\bar z} = 0$, these now reduce to
\begin{equation*}
\partial_{\bar z} A_\xi = 2 \pi \ii \sum_{m=1}^{N_1} U_m \delta(z - a_m), \qquad
\partial_{\bar z} A_\eta = 2 \pi \ii \sum_{n=1}^{N_2} V_n \delta(z - b_n).
\end{equation*}
Using the identity $\partial_{\bar z} z^{-1} = - 2 \pi \ii \delta(z)$ we deduce that a solution of the above is
\begin{equation} \label{A solution}
A_\xi = \L_\xi \coloneqq - U_0 - \sum_{m=1}^{N_1} \frac{U_m}{z - a_m}, \qquad
A_\eta = \L_\eta \coloneqq - V_0 - \sum_{n=1}^{N_2} \frac{V_n}{z - b_n}.
\end{equation}
These expressions coincide with those for the $d\xi$ and $d\eta$-components $U$ and $V$ of the Lax connection from \cite[(2) \& (6)]{ZakhMikh}.

Note that if we have $U_0 = d_\xi h h^{-1} $ and $V_0 = d_\eta h h^{-1}$ for some $h \in C^\infty(\Sigma, GL_N)$, cf. \cite[(5)]{ZakhMikh}, then we can set them both to zero in \eqref{A solution} using a gauge transformation with $g = h^{-1}$. This would have the effect of fixing the residual gauge symmetry down to the global transformations and the Lax connection \eqref{A solution} would then have no constant term, \emph{i.e.} it would take the form
\begin{equation*}
A_\xi = - \sum_{m=1}^{N_1} \frac{U_m}{z - a_m}, \qquad
A_\eta = - \sum_{n=1}^{N_2} \frac{V_n}{z - b_n}.
\end{equation*}
We will, however, keep the residual gauge symmetry for the remainder of this section, which will become the gauge symmetry in the $2$d action.

\subsection{Defect equations of motion}

We may also consider the variation of the action \eqref{defect 4d action} with respect to the $2$d defect fields $\phi_m$ and $\psi_n$. Consider the variation $\phi_m \mapsto e^{\epsilon \alpha_m} \phi_m$ for arbitrary $\alpha_m \in C^\infty(\Sigma, \gl_N)$ with $m = 1, \ldots, N_1$ and $\psi_n \mapsto e^{\epsilon \beta_n} \psi_n$ for arbitrary $\beta_n \in C^\infty(\Sigma, \gl_N)$ with $n = 1, \ldots, N_2$ in the action \eqref{defect 4d action}. This gives
\begin{align*}
&\delta_{(\alpha_m, \beta_n)} S\big( A, \{ \phi_m \}_{m=1}^{N_1}, \{ \psi_n \}_{n=1}^{N_2} \big) \coloneqq \frac{d}{d\epsilon} \bigg|_{\epsilon = 0} S\big( A, \{ e^{\epsilon \alpha_m} \phi_m \}_{m=1}^{N_1}, \{ e^{\epsilon \beta_n} \psi_n \}_{n=1}^{N_2} \big)\\
&\qquad\qquad\qquad\qquad\qquad = \sum_{m=1}^{N_1} \int_{\Sigma \times \{ a_m \}} \Tr\big( \alpha_m \big( d_\Sigma U_m - [U_m, \iota_{a_m}^\ast A] \big) \big) \wedge d\xi \notag\\
&\qquad\qquad\qquad\qquad\qquad\quad + \sum_{n=1}^{N_2} \int_{\Sigma \times \{ b_n \}} \Tr \big( \beta_n \big( d_\Sigma V_n - [V_n, \iota_{b_n}^\ast A] \big) \big) \wedge d\eta.
\end{align*}
Taking into account the solution \eqref{A solution} this then leads to the equations of motion
\begin{equation} \label{res ZC}
\partial_\eta U_m = - \bigg[ U_m, V_0 + \sum_{n=1}^{N_2} \frac{V_n}{a_m - b_n} \bigg], \qquad
\partial_\xi V_n = - \bigg[ V_n, U_0 + \sum_{m=1}^{N_1} \frac{U_m}{b_n - a_m} \bigg]
\end{equation}
for $m = 1, \ldots, N_1$ and $n = 1, \ldots, N_2$.
These coincide with \cite[(4)]{ZakhMikh} (noting that there is a sign mistake in \cite[(4)]{ZakhMikh}). Of course, the equations \eqref{res ZC} are nothing but the residues of the zero curvature equation \eqref{A flatness} at $a_m$ and $b_n$ respectively, taking into account the solution \eqref{A solution}.

\subsection{The Zakharov-Mikhailov action}

We now substitute the solution \eqref{A solution} for $A$, which we write as $\L = \L_\eta d\eta + \L_\xi d\xi$ since it correspond to the Lax connection, into the action \eqref{defect 4d action}.
The $4$d Chern-Simons action term becomes
\begin{equation} \label{CS part of action}
S_{4d}(\L) = - \frac{\ii}{4 \pi} \int_X z \, \Tr ( F(\L) \wedge F(\L) ) = - \frac{\ii}{2 \pi} \int_X z \, \Tr(\partial \L \wedge \bar\partial \L )
\end{equation}
where in the second equality we used the fact that $\L$ only has components along $d\eta$ and $d\xi$ which implies, in particular, that $\Tr(d \L \wedge \L \wedge \L ) = 0$.
Using the explicit form \eqref{A solution} of $\L$ we find
\begin{align*}
\Tr ( \partial \L \wedge \bar\partial \L ) &= 2 \pi \ii \sum_{m=1}^{N_1} \sum_{n=1}^{N_2} \frac{\Tr(U_m V_n) \big( \delta(z - b_n) - \delta(z - a_m) \big)}{(a_m - b_n)^2} dz \wedge d\bar z \wedge d\eta \wedge d\xi.
\end{align*}
Substituting this into \eqref{CS part of action} and performing the integral over $dz \wedge d\bar z$ by using the property \eqref{delta property} of the $\delta$-function we find
\begin{align} \label{4d CS term}
S_{4d}(\L) &= - \sum_{m=1}^{N_1} \sum_{n=1}^{N_2} \int_\Sigma \frac{\Tr (U_m V_n)}{a_m - b_n} d\eta \wedge d\xi.
\end{align}

On the other hand, substituting the solution \eqref{A solution} for $A$ into the two surface defect contributions to the action, namely \eqref{defect 2d action}, we obtain
\begin{align} \label{defect terms}
&S_{\rm defect}\big( A, \{ \phi_m \}_{m=1}^{N_1}, \{ \psi_n \}_{n=1}^{N_2} \big) = - \int_\Sigma \Tr \bigg( \sum_{n=1}^{N_1} \phi_n^{-1} (\partial_\eta - V_0) \phi_n U^{(0)}_n \notag\\
&\qquad\qquad\qquad - \sum_{n=1}^{N_2} \psi_n^{-1} (\partial_\xi - U_0) \psi_n V^{(0)}_n - 2 \sum_{m=1}^{N_1} \sum_{n=1}^{N_2} \frac{U_m V_n}{a_m - b_n} \bigg) d\eta \wedge d\xi.
\end{align}

Combining together \eqref{4d CS term} and \eqref{defect terms}, and recalling that $U_m = \phi_m U^{(0)}_m \phi_m^{-1}$ and $V_n = \psi_n V^{(0)}_n \psi_n^{-1}$, we arrive at the following $2$d action
\begin{align} \label{ZM action}
&S_{\rm 2d}\big( \{ \phi_n \}_{n=1}^{N_1}, \{ \psi_n \}_{n=1}^{N_2} \big) \notag\\
&\quad = - \int_\Sigma \Tr \bigg( \sum_{n=1}^{N_1} \phi_n^{-1} (\partial_\eta - V_0) \phi_n U^{(0)}_n - \sum_{n=1}^{N_2} \psi_n^{-1} (\partial_\xi - U_0) \psi_n V^{(0)}_n \notag\\
&\qquad\qquad\qquad\qquad\qquad\qquad\qquad - \sum_{m=1}^{N_1} \sum_{n=1}^{N_2} \frac{\phi_m U^{(0)}_m \phi_m^{-1} \psi_n V^{(0)}_n \psi_n^{-1}}{a_m - b_n} \bigg) d\eta \wedge d\xi.
\end{align}
This coincides with the Zakharov-Mikhailov action \cite[(10)]{ZakhMikh} up to an overall sign.

\subsection{Example}

The simplest non-trivial example of the Zakharov-Mikhailov action is obtained by taking $N_1 = N_2 = 1$.
In this case we only have two fields $\phi_1$ and $\psi_1$ which we denote simply as $\phi$ and $\psi$. Moving to a gauge where $V_0 = U_0 = 0$, as described in \S\ref{sec: Lax connection}, and choosing $U^{(0)} = - \Lambda$ and $V^{(0)} = \Lambda$ for some fixed constant matrix $\Lambda$, the action \eqref{ZM action} takes the simple form
\begin{equation*}
S_{\rm 2d}(\phi, \psi) = \int_\Sigma \Tr \bigg( \phi^{-1} \partial_\eta \phi \Lambda + \psi^{-1} \partial_\xi \psi \Lambda + \frac{1}{2\nu} \phi \Lambda \phi^{-1} \psi \Lambda \psi^{-1} \bigg) d\eta \wedge d\xi,
\end{equation*}
where we have introduced the coupling $2\nu \coloneqq a_1 - b_1$.

This action coincides with that of the so-called linear chiral model constructed in \cite[(3.20)]{Appadu:2017fff}. The latter can be seen as a generalisation to an arbitrary Lie algebra (here written only for $\gl_N$) of the model proposed by Faddeev and Reshetikhin in \cite{Faddeev:1985qu} as an ultralocal reduction of the $SU(2)$ principal chiral model. More precisely, the Faddeev-Reshetikhin model is defined by replacing the non-ultralocal Poisson bracket of the $SU(2)$ principal chiral model by an ultralocal one. However, the latter is degenerate and therefore the Faddeev-Reshetikhin model can only reproduce a reduction of the original principal chiral model dynamics, in which the Casimirs of the ultralocal Poisson bracket have been set to constants. In the next section we will derive the covariant Poisson algebra of the Lax connection \eqref{A solution} which in the present two-point case generalises the ultralocal algebra for the Lax matrix of the linear chiral model found in \cite[(3.5)]{Appadu:2017fff}.

\section{Covariant Poisson bracket and $r$-matrix for the $2$d theory} \label{sec: covariant}

In this section, we will rely heavily on the calculus in the variational bicomplex as presented in \cite{Dickey}. Informally, we introduce two differentials: $d$ is the ``horizontal'' differential, and acts as the usual exterior differential, while $\delta$ is the ``vertical'' differential that acts only with respect to the fields. We consider $(p,q)$-differential forms that have a vertical degree $p$ and a horizontal degree $q$. For instance, $\L = \L_\eta d\eta + L_\xi d\xi$ is a $(0,1)$-form, or a horizontal 1-form, and $\Omega^{(1)}$ below \eqref{omegaone} is a $(1,1)$-form. For details on how this is used in deriving the $r$-matrix structure of the covariant Poisson bracket of the Lax connection of a $2$d integrable field theory, or more generally an integrable hierarchy, we refer the reader to \cite{CS1,CS3}. 

We proceed in four steps: to begin with, we derive the multisymplectic form of the theory by considering the variation of its Lagrangian volume form, as established in \cite[(19.5.2)]{Dickey} and then used in \cite{CS1}. We can then define the covariant Poisson bracket of certain horizontal forms, called Hamiltonian, using the multisymplectic form. We then show that the Lax form associated with the Zakharov-Mikhailov theory is Hamiltonian and compute its covariant Poisson bracket structure \`a la Sklyanin \cite{Skly1, Skly2}, thus exhibiting its $r$-matrix structure. Finally, we construct the covariant Hamiltonian for the 2$d$ theory, which is the covariant analogue of the usual Hamiltonian obtained by performing the Legendre transformation with respect to both independent variables, and we interpret the zero-curvature equations as covariant Hamilton equations.

\subsection{The multisymplectic form}

Our starting point is the Lagrangian volume form associated with \eqref{ZM action}, where from now on we shall drop the inessential overall minus sign compared to \cite[(10)]{ZakhMikh}. However, throughout this section we shall work in the gauge where $U_0 = V_0 = 0$, so we start from
\begin{equation*} 
L_{\rm ZM} \coloneqq \Tr\bigg( \sum_{m=1}^{N_1} \phi_m^{-1} \partial_\eta \phi_m U^{(0)}_m - \sum_{n=1}^{N_2} \psi_n^{-1} \partial_\xi \psi_n V^{(0)}_n - \sum_{m=1}^{N_1} \sum_{n=1}^{N_2} \frac{U_m V_n}{a_m - b_n} \bigg) d\eta \wedge d\xi,
\end{equation*}
where we recall the notations $U_m = \phi_m U^{(0)}_m \phi_m^{-1}$ and $V_n=\psi_nV^{(0)}_n \psi_n^{-1}$. In particular, we have $\delta U_m = [\delta \phi_m \phi_m^{-1}, U_m]$ and $\delta V_n = [\delta \psi_n \psi_n^{-1}, V_n]$.
We also note the identities
\begin{align*}
&\delta \big( \Tr\big( \phi_m^{-1} \partial_\eta \phi_m U^{(0)}_m \big) d\eta \wedge d\xi \big)\\
&\qquad\qquad = \Tr\big( \! - \partial_\eta U_m \delta \phi_m \phi_m^{-1} \big) \wedge d\eta \wedge d\xi - d\Tr \big( \phi_m^{-1} \delta \phi_m U^{(0)}_m \wedge d\xi\big),\\
- \, &\delta \big( \Tr\big( \psi_n^{-1} \partial_\xi \psi_n V^{(0)}_n \big) d\eta \wedge d\xi \big)\\
&\qquad\qquad = \Tr\big( \partial_\xi V_n \delta \psi_n \psi_n^{-1} \big) \wedge d\eta \wedge d\xi - d\Tr \big( \psi_n^{-1} \delta \psi_n V^{(0)}_n \wedge d\eta \big).
\end{align*}
To show these we need, in particular, to use the fact that $\delta d = - d \delta$ along with the cyclicity of the trace. Combining the above we then find
\begin{align} \label{variation LZM}
\delta L_{\rm ZM}=& \Tr\Bigg( \!\! - \sum_{m=1}^{N_1} \bigg( \partial_\eta U_m + \sum_{n=1}^{N_2} \frac{[U_m,V_n]}{a_m - b_n} \bigg) \delta\phi_m \phi_m^{-1} \notag\\
&\qquad\qquad\qquad + \sum_{n=1}^{N_2} \bigg( \partial_\xi V_n + \sum_{m=1}^{N_1} \frac{[V_n,U_m]}{b_n - a_m} \bigg) \delta\psi_n \psi_n^{-1} \Bigg) d\eta \wedge d\xi\notag\\
&\quad -d \Tr \bigg(\sum_{m=1}^{N_1} \phi_m^{-1} \delta \phi_m U^{(0)}_m \wedge d\xi+\sum_{n=1}^{N_2} \psi_n^{-1} \delta \psi_n V^{(0)}_n \wedge d\eta \bigg)\,.
\end{align}

As expected, the first term reproduces the Euler-Lagrange equations in the form \eqref{res ZC}, recalling that we are working in the gauge where $U_0 = V_0 = 0$. On the other hand, the last term on the right hand side of \eqref{variation LZM} allows us to identify the form
\begin{equation}\label{omegaone}
\Omegaone=\sum_{m=1}^{N_1} \Tr(\phi_m^{-1} \delta \phi_m U^{(0)}_m) \wedge d\xi+\sum_{n=1}^{N_2} \Tr (\psi_n^{-1} \delta \psi_n V^{(0)}_n) \wedge d\eta\,,
\end{equation}
which in turn yields the multisymplectic form $\Omega \coloneqq \delta \Omegaone$ of the model as
\begin{equation}\label{omega}
\Omega=-\sum_{m=1}^{N_1} \Tr \big( \phi_m^{-1} \delta \phi_m \wedge \phi_m^{-1} \delta \phi_m U^{(0)}_m \big) \wedge d\xi-\sum_{n=1}^{N_2} \Tr \big( \psi_n^{-1} \delta \psi_n\wedge \psi_n^{-1} \delta \psi_n V^{(0)}_n \big) \wedge d\eta.
\end{equation}
The multisymplectic form $\Omega = \omega_{(\xi)} \wedge d\xi + \omega_{(\eta)} \wedge d\eta$ provides the covariant symplectic structure of a field theory. Its coefficients $\omega_{(\xi)}$ and $\omega_{(\eta)}$ contain the pull-back to the group of the Kostant-Kirillov forms for the orbits through $U^{(0)}_m$ and $V^{(0)}_n$ respectively.

\subsection{Covariant Poisson bracket of Hamiltonian $1$-forms}

We are now ready to define the covariant Poisson bracket $\cpb{~}{~}$ between certain horizontal forms called Hamiltonian. Specifically, a horizontal form $F$ is \emph{Hamiltonian} if there exists a vector field $X_F$ such that 
\begin{equation} \label{Ham vec field}
\delta F=\ip{X_F}{\Omega}\,,
\end{equation}
where $\ip{~}{~}$ denotes the interior product of a vector field with a form.
Let $F$ and $G$ be two Hamiltonian forms . We define their covariant Poisson bracket as follows
\begin{equation} \label{cov PB def}
\cpb{F}{G} \coloneqq (-1)^q \ip{X_F}{\delta G}=(-1)^q\ip{X_F}{\ip{X_G}{\Omega}}\,,
\end{equation}
where $q$ is the horizontal degree of $F$.
Notice that the vector field $X_F$ in \eqref{Ham vec field} will generally not be unique since $\Omega$ may have a non-trivial kernel. Nevertheless, the covariant Poisson bracket \eqref{cov PB def} is seen to be independent of the choice of vector fields $X_F$ and $X_G$ for both of the Hamiltonian forms $F$ and $G$. 
We remark that the covariant Poisson bracket is non-trivial and well defined when the Hamiltonian forms $F$ and $G$ are either both horizontal 1-forms, or one is a horizontal 1-form and the other one is a 0-form (\ie a function).

Our objective is to compute the covariant Poisson bracket \`a la Sklyanin for the Lax connection $\L = \L_\eta d\eta + \L_\xi d\xi$ corresponding to the solution \eqref{A solution} for $A$, in the gauge where $U_0=V_0 = 0$. Specifically, let $E_{ij}$ be the canonical basis for $\gl_N$ and write the Lax connection in this basis as
\begin{equation*}
\L(z) = \sum_{i,j=1}^N\L_{ij}(z) \,E_{ij},
\end{equation*}
where from now on we shall show the explicit dependence on the spectral parameter.
To compute the covariant Poisson brackets between any two components of the Lax connection, we first need to show that these are Hamiltonian $1$-forms.

For this we shall need the following useful identities. If $M$ is any $GL_N$-valued field with components $M_{ij}$, $i,j=1,\dots,N$ and $C$ is any non-dynamical matrix (meaning $\delta C=0$), then we have
\begin{subequations}
\begin{align}
\label{identity1}
&\ip{\sum_{k=1}^N M_{ik}\parder{}{M_{jk}}}{\Tr \left(M^{-1}\delta M\wedge M^{-1}\delta M C\right)}=\delta(MCM^{-1})_{ij}\,,\\
\label{identity2}
&\ip{\sum_{k=1}^N M_{ik}\parder{}{M_{jk}}}{\delta(MCM^{-1})_{kl}}=\delta_{jk}(MCM^{-1})_{il}-\delta_{il}(MCM^{-1})_{kj}\,.
\end{align}
\end{subequations}
In particular, we can use these with $M=\phi_n$, $C=U_n^{(0)}$ and $M=\psi_n$, $C=V_n^{(0)}$. Then a direct calculation shows that 
\be
X_{ij}(z) = \sum_{m=1}^{N_1}\sum_{\beta=1}^{N} \frac{\phi_{m,i\beta}}{z - a_m} \parder{}{\phi_{m,j\beta}} + \sum_{n=1}^{N_2}\sum_{\beta=1}^{N} \frac{\psi_{n,i\beta}}{z - b_n} \parder{}{\psi_{n,j\beta}}\,,
\ee
satisfies $\delta \L_{ij}(z)=\ip{X_{ij}(z)}{\Omega}$. Therefore all the components $\L_{ij}(z)$ for $i,j=1,\dots,N$ of the Lax connection are Hamiltonian $1$-forms, as required.

We shall write the covariant Poisson bracket of the Lax connection $\L$ using the standard tensorial notation $\L_1 \coloneqq \L\otimes\1$ and $\L_2 \coloneqq \1\otimes\L$ as
\begin{equation} \label{Lax cov PB}
\cpb{\L_1(z)}{\L_2(w)} \coloneqq \sum_{i,j=1}^N \cpb{\L_{ij}(z)}{\L_{kl}(w)}E_{ij}\otimes E_{kl}\,.
\end{equation}

\subsection{The $r$-matrix structure}

We now turn to the computation of the components on the right hand side of \eqref{Lax cov PB}. We have
\begin{align*}
\cpb{\L_{ij}(z)}{\L_{kl}(w)} &= -\ip{X_{ij}(z)}{\delta\L_{kl}(w)}\nonumber\\
&= \sum_{m=1}^{N_1}\frac{\delta_{jk}(U_m)_{il}-\delta_{il}(U_m)_{kj}}{(z-a_m)(w-a_m)} d\xi + \sum_{n=1}^{N_2}\frac{\delta_{jk}(V_n)_{il}-\delta_{il}(V_n)_{kj}}{(z-b_n)(w-b_n)} d\eta.
\end{align*}
Introducing the permutation operator $P_{12} \coloneqq \sum_{i,j=1}^NE_{ij}\otimes E_{ji}$ with the property
\begin{equation*}
\sum_{i,j=1}^N\left(\delta_{jk}M_{il}-\delta_{il}M_{kj}\right)E_{ij}\otimes E_{kl}=[M_1,P_{12}]=-[M_2,P_{12}]\,,
\end{equation*}
for any $\gl_N$-valued field $M$ with components $M_{ij}$ for $i,j = 1, \ldots, N$, and noting that for any distinct $z, w, a \in \CC$ we have the identity
\begin{equation} \label{equality}
\frac{1}{(z-a)(w-a)}=\frac{1}{w-z}\left(\frac{1}{z-a}-\frac{1}{w-a}\right),
\end{equation}
we may rewrite the covariant Poisson bracket \eqref{Lax cov PB} as
\begin{align*}
\cpb{\L_1(z)}{\L_2(w)}&=\sum_{m=1}^{N_1}\frac{[(U_m)_{1},P_{12}]}{(z-a_m)(w-a_m)} d\xi +\sum_{n=1}^{N_2}\frac{[(V_n)_{1},P_{12}]}{(z-b_n)(w-b_n)} d\eta\nonumber\\
&= \bigg[\frac{P_{12}}{z-w},\L_1(z)+\L_2(w)\bigg]\,.
\end{align*}
In other words, we have the announced result that the Lax connection satisfies the following Poisson algebra
\begin{equation*}
\cpb{\L_1(z)}{\L_2(w)}=\big[r_{12}(z-w),\L_1(z)+\L_2(w)\big]\,,
\end{equation*}
with respect to the {\it covariant} Poisson bracket $\cpb{~}{~}$, where $r_{12}(z) \coloneqq P_{12}/z$ is the rational $r$-matrix. The fact that we have been working with the Lie algebra $\gl_N$ was convenient for writing the $GL_N$-valued fields $\phi_n$ and $\psi_n$ in components. However, the above derivation can be adapted to hold more generally for any semisimple Lie algebra, working in a basis of the latter.

\subsection{The covariant Hamiltonian}

Following \cite[Lemma 19.5.9]{Dickey}, the covariant Hamiltonian related to $L_{\rm ZM}$ is found to be equal to
\begin{align*}
H_{\rm ZM} &\coloneqq -L_{\rm ZM} +\sum_{m=1}^{N_1}\Tr\big( \phi_m^{-1}\partial_\eta\phi_m U_m^{(0)} \big)\, d\eta \wedge d \xi - \sum_{n=1}^{N_2}\Tr\big( \psi_n^{-1}\partial_\xi\psi_n V_n^{(0)} \big)\, d\eta \wedge d \xi \nonumber\\
&\, = \sum_{m=1}^{N_1} \sum_{n=1}^{N_2} \Tr\frac{U_m V_n}{a_m - b_n} \, d\eta \wedge d \xi\,.
\end{align*}
This can be reexpressed directly in terms of the Lax connection as 
\begin{equation} \label{Ham ZM}
H_{\rm ZM}=\sum_{m=1}^{N_1} \sum_{n=1}^{N_2} \res_{z=a_m} \res_{w=b_n}\Tr\frac{\L(z)\wedge \L(w)}{z-w}\,.
\end{equation}
This is a rather remarkable formula extending to the present covariant context the familiar formula ``$H=\Tr L^2$'' for extracting a Hamiltonian from a Lax matrix in many finite-dimensional integrable systems, such as the Gaudin model or Calogero-Moser system. In fact, the formula \eqref{Ham ZM} is very reminiscent of the expression for the Hamiltonian in non-ultralocal integrable field theories described by Gaudin models associated with affine Kac-Moody algebras \cite{Vicedo:2017cge, Delduc:2019bcl}.

\subsection{Flatness of the Lax connection as a covariant Hamilton equation}

It was shown for the first time in \cite{CS1} for certain $2$d integrable field theories (nonlinear Schr\"odinger, sine-Gordon, modified Korteweg-de Vries) that the zero curvature equation \eqref{A flatness} is a covariant Hamilton equation for the Lax connection $\L$ associated with the density of the covariant Hamiltonian. By this we mean that, if we define the ``covariant flow'' of $\L$ by
\begin{equation*}
d\L(z)=\cpb{h_{\rm ZM}}{\L(z)}d\eta\wedge d\xi\,,~~\text{where}~~H_{ZM}=h_{ZM}d\eta\wedge d\xi\,,
\end{equation*}
in analogy to what one would do in the traditional Hamiltonian formalism, then since we have
\begin{equation} \label{covariant_Hamilton}
\cpb{h_{\rm ZM}}{\L(z)}d\eta\wedge d\xi=-\L(z)\wedge \L(z)\,,
\end{equation}
we can conclude that $d\L(z)+\L(z)\wedge \L(z)=0$. The main steps in the derivation of the crucial equality \eqref{covariant_Hamilton} are as follows. First, we have by definition
\begin{equation}
\label{cpb_comp}
\cpb{h_{\rm ZM}}{\L(z)}=\sum_{i,j=1}^N \ip{X_{ij}(z)}{\delta h_{\rm ZM}}  E_{ij}\,.
\end{equation}
Second, we find
\begin{align*}
\ip{X_{ij}(z)}{\delta h_{\rm ZM}} &= \bigg(\sum_{m=1}^{N_1}\sum_{\beta=1}^{N} \frac{\phi_{m,i\beta}}{z - a_m} \parder{}{\phi_{m,j\beta}} + \sum_{n=1}^{N_2}\sum_{\beta=1}^{N} \frac{\psi_{n,i\beta}}{z - b_n} \parder{}{\psi_{n,j\beta}}\bigg)\nonumber\\
&\qquad \lrcorner\bigg(\sum_{p=1}^{N_1}\sum_{q=1}^{N_2}\sum_{k,l=1}^N\frac{(\delta U_p)_{kl}(V_q)_{lk}+(U_p)_{lk}(\delta V_q)_{kl}}{a_p-b_q}  \bigg)\nonumber\\
&= \sum_{m=1}^{N_1}\sum_{q=1}^{N_2}\sum_{k,l=1}^N\frac{\left(\delta_{jk}(U_m)_{il}-\delta_{il}(U_m)_{kj}\right)(V_q)_{lk}}{(z-a_m)(a_m-b_q)}\nonumber\\
&\qquad + \sum_{n=1}^{N_2}\sum_{p=1}^{N_1}\sum_{k,l=1}^N\frac{(U_p)_{lk}\left(\delta_{jk}(V_n)_{il}-\delta_{il}(V_n)_{kj}\right)}{(z-b_n)(a_p-b_n)}\nonumber\\
&= \sum_{m=1}^{N_1}\sum_{n=1}^{N_2}\frac{\left([U_m,V_n] \right)_{ij}}{(z-a_m)(z-b_n)}\,,
\end{align*}
 where we have used the identity \eqref{identity2} in the second equality and \eqref{equality} in the last equality. Substituting the above into \eqref{cpb_comp} we obtain \eqref{covariant_Hamilton}.

\section{Conclusion and outlook}

In this paper we derived the Zakharov-Mikhailov $2$d action from the $4$d Chern-Simons action in the presence of certain surface defects. At the $2$d level, the covariant Poisson algebra of the Lax connection was shown to possess a classical $r$-matrix structure of rational type, thereby recasting the pioneering results of Sklyanin \cite{Skly1,Skly2} into a covariant Hamiltonian context. So far, this had only been shown for the sine-Gordon model \cite{CS1} and the entire AKNS hierarchy \cite{CS2,CS3}. 
There are a number of tantalising questions and possible further directions following this work.

\medskip

Some of the models (e.g. deformed Gross-Neveu models) considered in the series of papers \cite{By1,By2,By3,By4} seem to be cousins of the models of Zakharov-Mikhailov type studied here. It would be natural to expect that the covariant Poisson algebra of the Lax connection also holds for these models. Whether this could be achieved by relating them to the present Zakharov-Mikhailov construction is an interesting problem. The extension of the covariant Poisson algebra structure to an entire hierarchy, as obtained in \cite{CS3}, is based on the notions of Hamiltonian multiform and multi-time Poisson bracket introduced in \cite{CS2}. In turn, these are based on the idea of Lagrangian multiforms \cite{LN} which provide a generalised variational principle that is able to capture the integrability properties of classical field theories. The Zakharov-Mikhailov action was analysed from this point of view and embedded into a Lagrangian multiform in \cite{SNC}. It is an intriguing problem to understand how such a multiform could effectively arise from a higher dimensional theory, in parallel to the present situation where a single $2$d Lagrangian is derived from a $4$d one. 

The Poisson algebra of the Lax matrix of a non-ultralocal $2$d integrable field theory was derived in \cite{Vicedo:2019dej} by performing a Hamiltonian analysis of the $4$d Chern-Simons action for a general $1$-form $\omega$. It would be interesting to similarly derive the covariant Poisson algebra of the Lax connection in the present ultralocal setting for which $\omega = dz$. This would involve performing a covariant Hamiltonian analysis of the $4$d action \eqref{defect 4d action} in order to rederive the covariant Poisson bracket obtained in \S\ref{sec: covariant} directly from the $4$d Chern-Simons theory.

We showed that the gauge transformations in the Zakharov-Mikhailov action arose as special types of gauge transformations in the $4$d Chern-Simons theory for which the gauge transformation parameter $g \in C^\infty(\Sigma, G)$ is independent of the spectral parameter. This is the crudest way of ensuring \eqref{A bar z gauge} but we believe that a more appropriate condition would be to require that $g$ is (sectionally) holomorphic in order to make a connection with the theory of dressing transformations \cite{Zakharov:1979zz}. In other words, it would be interesting to understand if dressing transformations in the $2$d integrable field theory can also be understood as arising from gauge transformations in $4$d Chern-Simons theory by allowing $g \in C^\infty(X, G)$ to depend also on $\CP$ as long as the pole structure of the Lax connection \eqref{A solution} remains unchanged under such gauge transformations. 

In the ultralocal setting considered in the present paper, the regularised $4$d Chern-Simons action is easily seen to be gauge invariant. Therefore any defect terms added to the action, as in \eqref{defect 4d action}, should be gauge invariant themselves. By contrast, in the non-ultralocal setting one needs to impose boundary conditions on the bulk field $A$ at the disorder defects, which are located at the poles of $\omega$ \cite{Costello:2019tri}. This is necessary in order to ensure that the action is gauge invariant \cite{Benini:2020skc}. Alternatively, the gauge invariance can be ensured by introducing new fields living on the surface defects, called the edge modes, and coupling these to the bulk field $A$ \cite{Benini:2020skc}. It would be very interesting to explore the possibility of combining these two approaches by adding further gauge invariant defect terms to the $4$d Chern-Simons action with edge modes. This would have the interesting effect of coupling, in the sense of \cite{Delduc:2018hty, Delduc:2019bcl}, ultralocal integrable field theories to a non-ultralocal one.

\section*{Acknowledgements}
V.C. and M.S want to acknowledge regular stimulating discussions with F. Nijhoff.

\section*{Conflict of interest}
On behalf of all authors, the corresponding author states that there is no conflict of interest.

\end{document}